\documentclass[aps,prd,draft,showpacs]{revtex4}

\begin{document}

\title{Gauge Parameter Independence of the Higgs Pole Mass in the Multi-Higgs model}
\author{Chungku Kim}
\date{\today}

\begin{abstract}
We have investigated the gauge parameter dependence of the Higgs
pole masses in the multi-Higgs model in two different types of the gauge
fixing conditions. It turns out that the Higgs pole masses of the multi-Higgs
model when expressed in terms of the Lagrangian parameters, are independent
of the gauge parameter for both cases.
\end{abstract}


\maketitle

\email{kimck@kmu.ac.kr}
\affiliation{Department of Physics, Keimyung University, Daegu 704-701, Korea}

\section{\bigskip INTRODUCTION}
Recently, gauge field theory which contains several Higgs field became of
interest to solve the questions of the neutrino mass, the matter-antimatter
in the universe and the nature of the dark matter\cite{Multi}. In obtaining
the physical results from these models, the pole mass plays an important
role in the process where the characteristic scale is close to the mass shell%
\cite{Pole}. It was shown that the pole mass is infrared finite\cite
{Infrared} and invariant under the renormalization group\cite{RG}. In case
of the quantized gauge field theory, we need to add a gauge-fixing terms\cite
{Gauge} which contain a gauge parameters and the physical quantities such as
the pole mass should be independant of these gauge parameters. In case of the
standard model which have single Higgs boson, the gauge independence of the Higgs
pole mass for the Higgs boson was shown in Ref.\cite{parameter}. In this paper,
we will investigate the gauge parameter dependence of the Higgs pole masses in
multi-Higgs models in two different types of the gauge fixing conditions.
\section{\protect GAUGE DEPENDENCE OF THE POLE MASS IN MULTI-HIGGS MODEL}
In the broken symmetry phase of the multi-Higgs model, the Higgs pole masses
$M_{i}^{2}$ are determined from the condition\cite{Quiros} 
\begin{equation}
\left[ \det \Pi _{pq}(p^{2})\right] _{p^{2}=-M_{i}^{2}}=0,
\end{equation}
where the renormalized inverse Higgs propagator $\Pi _{pq}(p^{2})$ is obtained
from the renormalized effective action in the broken symmetry phase $\Gamma $
as 
\begin{equation}
\Pi _{pq}(p^{2})=\left[ \frac{\delta ^{2}\Gamma }{\delta h_{p}\delta h_{q}}%
\right] _{h=0}.
\end{equation}
where $h$ is the classical field corresponding to the Higgs field $H$ of the
Lagrangian density.
Then the gauge parameter $\xi ^{\alpha }$, corresponding to $\alpha -th$ gauge boson,
dependence of the Higgs pole mass can be
obtained by taking $\xi ^{\alpha }\frac{\partial }{\partial \xi ^{\alpha }}$
to Eq.(1) to obtain as
\begin{eqnarray}
0 &=&\xi ^{\alpha }\frac{\partial }{\partial \xi ^{\alpha }}\left[ \det \Pi
(p^{2})\right] _{p^{2}=-M_{i}^{2}}=-(\xi ^{\alpha }\frac{\partial M_{i}^{2}}{%
\partial \xi ^{\alpha }})\left[ \frac{\delta }{\delta p^{2}}\det \Pi
(p^{2})\right] _{p^{2}=-M_{i}^{2}}+\left[ \xi ^{\alpha }\frac{\partial }{%
\partial \xi ^{\alpha }}\det \Pi (p^{2})\right] _{p^{2}=-M_{i}^{2}} 
\nonumber \\
&=&-(\xi ^{\alpha }\frac{\partial M_{i}^{2}}{\partial \xi ^{\alpha }})\left[ 
\frac{\delta }{\delta p^{2}}\det \Pi (p^{2})\right]
_{p^{2}=-M_{i}^{2}}+\left[ Tr\{\Gamma ^{-1}(p^{2})\xi ^{\alpha }\frac{%
\partial }{\partial \xi ^{\alpha }}\Pi (p^{2})\}\det \Pi (p^{2})\right]
_{p^{2}=-M_{i}^{2}}  \nonumber \\
&=&-(\xi ^{\alpha }\frac{\partial M_{i}^{2}}{\partial \xi ^{\alpha }})\left[ 
\frac{\delta }{\delta p^{2}}\det \Pi (p^{2})\right]
_{p^{2}=-M_{i}^{2}}+\left[ Tr\{adj(\Pi (p^{2})\xi ^{\alpha }\frac{\partial }{%
\partial \xi ^{\alpha }}\Pi (p^{2})\}\right] _{p^{2}=-M_{i}^{2}},
\end{eqnarray}
where $adj(\Pi (p^{2}))$ is the adjoint of \ $\Pi (p^{2})$ given as 
\begin{equation}
adj\{\Pi (p^{2})\}=\Pi ^{-1}(p^{2})\det \Pi (p^{2}).
\end{equation}
Note that although $\left[ \det \Pi _{pq}(p^{2})\right] _{p^{2}=-M_{i}^{2}}$
vanishes, $adj\{\Pi (p^{2})\}$ is nonzero when $p^{2}=-M_{i}^{2}$.
It follows that if $\xi ^{\alpha }\frac{\partial }{\partial \xi ^{\alpha }}%
\Pi (p^{2})$ have a form 
\begin{equation}
\xi ^{\alpha }\frac{\partial }{\partial \xi ^{\alpha }}\Pi (p^{2})=A_{1}\Pi
+\Pi A_{2},
\end{equation}
for some matrices $A_{1}$and $A_{2}$,\ then $\left[ Tr\{adj(\Pi (p^{2})\xi
^{\alpha }\frac{\partial }{\partial \xi ^{\alpha }}\Pi (p^{2})\}\right]
_{p^{2}=-M_{i}^{2}}$ term of the last line of Eq.(3) is proportional to $%
\left[ Tr\{\det \Pi (p^{2})\}\right] _{p^{2}=-M_{i}^{2}}$ and vanishes due
to Eq.(1). \ As a result, we obtain 
\begin{equation}
\xi ^{\alpha }\frac{\partial M_{i}^{2}}{\partial \xi ^{\alpha }}=0,
\end{equation}
if Eq.(5) is satisfied.

In order to test whether Eq.(5) is satisfied or not in given gauge fixing
term, let us start from the Nielsen identity\cite{Nielsen} for the gauge
parameter $\xi ^{\alpha }$ in the symmetric phase of \ the multi-Higgs model
given as\cite{DelCima} 
\begin{equation}
\xi ^{\alpha }\frac{\partial \Gamma ^{SYM}(\phi )}{\partial \xi ^{\alpha }}%
+C_{i}^{\alpha }(\phi )\frac{\delta \Gamma ^{SYM}}{\delta \phi _{i}}=0\text{
(no sum on }\alpha \text{),}
\end{equation}
where $\phi_i$ is the classical field corresponding to the scalar field $\Phi_i$
of the Lagrangian density and $\Gamma ^{SYM}(\phi )$ is the effective action
in the symmetric phase and the index $i$ runs both the internal and the space-time
variables. In the broken symmetry phase, the gauge fixing can be divided into two
classes which we call class I and class II type respectively. In case of the class
I type of gauge fixing, the gauge fixing term in the broken symmetry phase
$L_{g.f.}$ is obtained from that of the symmetric phase $L_{g.f.}^{SYM}$ as 
\begin{equation}
L_{g.f.}(H_{i},v_{i})=\left[ L_{g.f.}^{SYM}(\Phi _{i})\right] _{\Phi
_{i}=H_{i}+v_{i}},
\end{equation}
where $v_{i}$ is the vacuum expectation value(VEV) for the scalar field $\Phi
_{i}$ and $H_{i}$ is the corresponding Higgs field. The typical case is the
non-linear $\overline{R_{\xi }}$ gauge\cite{Bar} and the Feynman gauge also
belongs to this class. 
In case of the class II type of the gauge fixing, the gauge fixing term in the
broken symmetry phase do not have a corresponding term in the symmetric phase
as in Eq.(8) and is obtained by identifying the gauge parameter $u_{i}$ of
the symmetric phase with the VEV in the broken symmetric phase as
\begin{equation}
L_{g.f.}(H_{i},v_{i})=\left[ L_{g.f.}^{SYM}(\Phi _{i},u_{i})\right]
_{u_{i}=v_{i},\Phi _{i}=H_{i}+v_{i}}.
\end{equation}
The typical case is the $R_{\xi }$ gauge fixing scheme of the form 
\begin{equation}
L_{g.f.}=\frac{1}{2\xi ^{\alpha }}(\partial _{\mu }G_{\mu }^{\alpha
}-gv_{i}T_{ij}^{\alpha }\Phi _{j})^{2}.
\end{equation}
in the broken symmetry phase which do not have a corresponding term in the
symmetric phase satisfying Eq.(8). In order to obtain this gauge fixing term
in the broken symmetry phase, we choose a gauge fixing term in the symmetric
phase of the form 
\begin{equation}
L_{g.f.}^{SYM}(\Phi _{i},u_{i})=\frac{1}{2\xi ^{\alpha }}(\partial _{\mu
}G_{\mu }^{\alpha }-gu_{i}T_{ij}^{\alpha }\Phi _{j})^{2},
\end{equation}
where $u_{i}$ plays a role of the gauge parameter in the symmetric phase\cite{u}.
Then, in the broken symmetry phase, we should substitute $v$ for$\ u$ to obtain
Eq.(9). 
Since $u_{i}$ is a gauge parameter in the symmetric phase, it also satisfies
a Nielsen identity of the form\cite{u} 
\begin{equation}
u_{i}\frac{\partial \Gamma ^{SYM}(\phi )}{\partial u_{i}}+C_{ij}(\phi )\frac{%
\delta \Gamma ^{SYM}}{\delta \phi _{j}}=0\text{ (no sum on }\alpha \text{).}
\end{equation}

Now, let us consider the gauge dependence of the VEVs $v_{p}$ corresponding to
p-th Higgs boson which satisfy 
\begin{equation}
\left[ \frac{\delta \Gamma ^{SYM}}{\delta \phi _{p}}\right] _{\phi
_{p}=v_{p}}=0.
\end{equation}
In case of the class I type of gauge fixing, by taking $\xi ^{\alpha }\frac{\delta }
{\delta \xi^{\alpha }}$ to Eq.(13) and by using Eq.(7), we obtain 
\begin{eqnarray}
0 &=&\xi ^{\alpha }\frac{\partial }{\partial \xi ^{\alpha }}\left[ \frac{%
\delta \Gamma ^{SYM}}{\delta \phi _{p}}\right] _{\phi _{p}=v_{p}}=\xi
^{\alpha }\frac{\partial v_{j}}{\partial \xi ^{\alpha }}\left[ \frac{\delta
^{2}\Gamma ^{SYM}}{\delta \phi _{p}\delta \phi _{j}}\right] _{\phi
_{p}=v_{p}}+\left[ \xi ^{\alpha }\frac{\partial }{\partial \xi ^{\alpha }}%
\frac{\delta \Gamma ^{SYM}}{\delta \phi _{p}}\right] _{\phi _{p}=v_{p}} 
\nonumber \\
&=&\left[ \xi ^{\alpha }\frac{\partial v_{j}}{\partial \xi ^{\alpha }}\frac{%
\delta ^{2}\Gamma ^{SYM}}{\delta \phi _{p}\delta \phi _{j}}-\frac{\delta
C_{j}^{\alpha }}{\delta \phi _{p}}\frac{\delta \Gamma ^{SYM}}{\delta \phi
_{j}}-C_{j}^{\alpha }\frac{\delta ^{2}\Gamma ^{SYM}}{\delta \phi _{p}\delta
\phi _{j}}\right] _{\phi _{p}=v_{p}}.
\end{eqnarray}
The second term of the last line of this equation vanishes due to Eq.(13) and
we obtain
\begin{equation}
\xi ^{\alpha }\frac{\partial v_{j}}{\partial \xi ^{\alpha }}=C_{j}^{\alpha
}(v).
\end{equation}
In case of the class II type of gauge fixing, since $u_{i}$ should be identified as
a VEV $v_{i}$ in the broken symmetry phase, Eq.(8) should be modified as 
\begin{equation}
\left[ \frac{\delta \Gamma ^{SYM}}{\delta \phi _{p}}\right] _{\phi
_{p}=v_{p},u_{i}=v_{i}}=0.
\end{equation}
This modification of the VEV causes the anomalous behavior of the gamma
function of the VEV\cite{gamma1}\cite{gamma2}. By taking $\xi ^{\alpha }%
\frac{\delta }{\delta \xi ^{\alpha }}$ to the Eq.(16) and by using Eqs.(7)
and (12), we obtain 
\begin{eqnarray}
0&=&\xi ^{\alpha }\frac{\partial }{\partial \xi ^{\alpha }}\left[ \frac{%
\delta \Gamma ^{SYM}}{\delta \phi _{p}}\right] _{\phi _{p}=u_{p}=v_{p}}
 \nonumber \\ &=& \xi
^{\alpha }\frac{\partial v_{i}}{\partial \xi ^{\alpha }}\left[ \frac{\delta
^{2}\Gamma ^{SYM}}{\delta \phi _{p}\delta \phi _{i}}+\frac{\partial }{%
\partial u_{i}}\frac{\delta \Gamma ^{SYM}}{\delta \phi _{p}}\right] _{\phi
_{p}=v_{p},u_{i}=v_{i}}+\left[ \xi ^{\alpha }\frac{\partial }{\partial \xi
^{\alpha }}\frac{\delta \Gamma ^{SYM}}{\delta \phi _{p}}\right] _{\phi
_{p}=v_{p},u_{i}=v_{i}}  \nonumber \\
&=&\xi ^{\alpha }\frac{\partial v_{i}}{\partial \xi ^{\alpha }}\left[ \frac{%
\delta ^{2}\Gamma ^{SYM}}{\delta \phi _{p}\delta \phi _{i}}-\frac{1}{u_{i}}%
\frac{\delta C_{ik}}{\delta \phi _{p}}\frac{\delta \Gamma ^{SYM}}{\delta
\phi _{k}}-\frac{1}{u_{i}}C_{ik}\frac{\delta ^{2}\Gamma ^{SYM}}{\delta \phi
_{p}\delta \phi _{k}}\right] _{\phi _{p}=v_{p},u_{i}=v_{i}}\nonumber \\
& &-\left[ \frac{%
\delta C_{i}^{\alpha }(\phi )}{\delta \phi _{p}}\frac{\delta \Gamma ^{SYM}}{%
\delta \phi _{i}}+C_{i}^{\alpha }(\phi )\frac{\delta ^{2}\Gamma ^{SYM}}{%
\delta \phi _{p}\delta \phi _{i}}+\right] _{\phi _{p}=v_{p},u_{i}=v_{i}}.
\end{eqnarray}
Then, applying Eq.(16) to last line of above equation, we obtain 
\begin{equation}
\xi ^{\alpha }\frac{\partial v_{i}}{\partial \xi ^{\alpha }}-\xi ^{\alpha }%
\frac{\partial v_{k}}{\partial \xi ^{\alpha }}\left[ \frac{1}{u_{k}}%
C_{ki}(\phi )\right] _{\phi _{p}=v_{p},u_{i}=v_{i}}=\left[ C_{i}^{\alpha
}(\phi )\right] _{\phi _{p}=v_{p},u_{i}=v_{i}}.
\end{equation}
Having obtained the gauge dependence of the VEVs in case of the \ class I
and II type of gauge fixing given in Eqs.(15) and (18), let us consider the
gauge dependence of the inverse Higgs propagator in the broken symmetry phase.
In case of the class I type of gauge fixing, the inverse Higgs propagator in
the broken symmetry phase is gven by 
\begin{equation}
\Pi _{pq}(p^{2})=\left[ \frac{\delta ^{2}\Gamma }{\delta h_{p}\delta h_{q}}%
\right] _{h_{i}=0}=\left[ \frac{\delta ^{2}\Gamma ^{SYM}(\phi )}{\delta \phi
_{p}\delta \phi _{q}}\right] _{\phi _{i}=v_{i}},
\end{equation}
from which we obtain 
\begin{eqnarray}
&&\xi ^{\alpha }\frac{\delta }{\delta \xi ^{\alpha }}\Pi _{pq}(p^{2}) 
\nonumber \\ &=&\xi
^{\alpha }\frac{\partial v_{i}}{\partial \xi ^{\alpha }}\left[ \frac{\delta
^{3}\Gamma ^{SYM}}{\delta \phi _{i}\delta \phi _{p}\delta \phi _{q}}\right]
_{\phi _{p}=v_{p}}-\left[ \frac{\delta ^{2}}{\delta \phi _{p}\delta \phi _{q}%
}\{C_{i}^{\alpha }(\phi )\frac{\delta \Gamma ^{SYM}}{\delta \phi _{i}}%
\}\right] _{\phi _{p}=v_{p}}=\xi ^{\alpha }\frac{\partial v_{i}}{\partial
\xi ^{\alpha }}\left[ \frac{\delta ^{3}\Gamma ^{SYM}}{\delta \phi _{i}\delta
\phi _{p}\delta \phi _{q}}\right] _{\phi _{p}=v_{p}}  \nonumber \\
&&-\left[ \frac{\delta ^{2}C_{i}^{\alpha }(\phi )}{\delta \phi _{p}\delta
h_{q}}\frac{\delta \Gamma ^{SYM}}{\delta \phi _{k}}+\frac{\delta
C_{i}^{\alpha }(\phi )}{\delta \phi _{p}}\frac{\delta ^{2}\Gamma ^{SYM}}{%
\delta \phi _{i}\delta \phi _{q}}+\frac{\delta C_{i}^{\alpha }(\phi )}{%
\delta \phi _{q}}\frac{\delta ^{2}\Gamma ^{SYM}}{\delta \phi _{i}\delta \phi
_{p}}+C_{i}^{\alpha }(\phi )\frac{\delta ^{3}\Gamma ^{SYM}}{\delta \phi
_{i}\delta \phi _{p}\delta \phi _{q}}\right] _{\phi _{p}=v_{p}}  \nonumber \\
&=&-\left[ \frac{\delta C_{i}^{\alpha }(\phi )}{\delta \phi _{p}}\right]
_{\phi _{p}=v_{p}}\Pi _{qi}(p^{2})-\left[ \frac{\delta C_{i}^{\alpha }(\phi )%
}{\delta \phi _{q}}\right] _{\phi _{p}=v_{p}}\Pi _{pi}(p^{2}),
\end{eqnarray}
where we have used Eqs.(7) and (15). Now, consider the case of the class II
type of gauge fixing where the inverse Higgs propagator in the broken symmetry
phase is gven by 
\begin{equation}
\Pi _{pq}(p^{2})=\left[ \frac{\delta ^{2}\Gamma }{\delta h_{p}\delta h_{q}}%
\right] _{h_{i}=0}=\left[ \frac{\delta ^{2}\Gamma ^{SYM}(\phi )}{\delta \phi
_{p}\delta \phi _{q}}\right] _{\phi _{p}=v_{p},u_{i}=v_{i}}.
\end{equation}
Then we obtain 
\begin{eqnarray}
&&\xi ^{\alpha }\frac{\partial }{\partial \xi ^{\alpha }}\Pi _{pq}(p^{2})\nonumber \\
&=&\xi ^{\alpha }\frac{\partial v_{i}}{\partial \xi ^{\alpha }}\left[ \frac{%
\delta ^{3}\Gamma ^{SYM}}{\delta \phi _{i}\delta \phi _{p}\delta \phi _{q}}+%
\frac{\partial }{\partial u_{i}}\frac{\delta ^{2}\Gamma ^{SYM}}{\delta \phi
_{p}\delta \phi _{q}}\right] _{\phi _{p}=v_{p},u_{i}=v_{i}}-\left[ \frac{%
\delta ^{2}}{\delta \phi _{p}\delta \phi _{q}}\{C_{i}^{\alpha }(\phi )\frac{%
\delta \Gamma ^{SYM}}{\delta \phi _{i}}\}\right] _{\phi
_{p}=v_{p},u_{i}=v_{i}}  \nonumber \\
&=&\xi ^{\alpha }\frac{\partial v_{i}}{\partial \xi ^{\alpha }}\left[ \frac{%
\delta ^{3}\Gamma ^{SYM}}{\delta \phi _{i}\delta \phi _{p}\delta \phi _{q}}%
\right] _{\phi _{p}=v_{p},u_{i}=v_{i}}  \nonumber \\
&&-\xi ^{\alpha }\frac{\partial v_{k}}{\partial \xi ^{\alpha }}\left[ \frac{1%
}{u_{k}}\{\frac{\delta ^{2}C_{ki}}{\delta \phi _{p}\delta \phi _{q}}\frac{%
\delta \Gamma ^{SYM}}{\delta \phi _{i}}+\frac{\delta C_{ki}}{\delta \phi _{p}%
}\frac{\delta ^{2}\Gamma ^{SYM}}{\delta \phi _{q}\delta \phi _{i}}+\frac{%
\delta C_{ki}}{\delta \phi _{q}}\frac{\delta ^{2}\Gamma ^{SYM}}{\delta \phi
_{i}\delta \phi _{p}}+C_{ki}\frac{\delta ^{3}\Gamma ^{SYM}}{\delta \phi
_{i}\delta \phi _{p}\delta \phi _{q}}\}\right] _{\phi _{p}=v_{p},u_{i}=v_{i}}
\nonumber \\
&&-\left[ \frac{\delta ^{2}C_{i}^{\alpha }(\phi )}{\delta \phi _{p}\delta
\phi _{q}}\frac{\delta \Gamma ^{SYM}}{\delta \phi _{i}}+\frac{\delta
C_{i}^{\alpha }(\phi )}{\delta \phi _{p}}\frac{\delta ^{2}\Gamma ^{SYM}}{%
\delta \phi _{i}\delta \phi _{p}}+\frac{\delta C_{i}^{\alpha }(\phi )}{%
\delta \phi _{q}}\frac{\delta ^{2}\Gamma ^{SYM}}{\delta \phi _{i}\delta \phi
_{p}}+C_{i}^{\alpha }(\phi )\frac{\delta ^{3}\Gamma ^{SYM}}{\delta \phi
_{i}\delta \phi _{p}\delta \phi _{q}}\right] _{\phi _{p}=v_{p},u_{i}=v_{i}} 
\nonumber \\
&=&-\left[ \frac{\xi ^{\alpha }}{v_{k}}\frac{\partial v_{k}}{\partial \xi
^{\alpha }}\frac{\delta C_{ki}}{\delta \phi _{p}}+\frac{\delta C_{i}^{\alpha
}(\phi )}{\delta \phi _{p}}\right] _{\phi _{p}=v_{p},u_{i}=v_{i}}\Pi
_{qi}(p^{2})-\left[ \frac{\xi ^{\alpha }}{v_{k}}\frac{\partial v_{k}}{%
\partial \xi ^{\alpha }}\frac{\delta C_{ki}}{\delta \phi _{q}}+\frac{\delta
C_{i}^{\alpha }(\phi )}{\delta \phi _{p}}\right] _{\phi
_{p}=v_{p},u_{i}=v_{i}}\Pi _{pi}(p^{2}),\nonumber \\
\end{eqnarray}
where we have used Eqs.(7),(12) and (18). From Eqs.(20) and (22), we can see
that both the class I and II gauge fixing satisfy Eq.(5) with
\begin{equation}
A_{1ij}^{T}=A_{2ij}=-\left[ \frac{\delta C_{i}^{\alpha }(\phi )}{\delta \phi
_{j}}\right] _{\phi _{p}=v_{p}},
\end{equation}
in case of class I type of gauge fixing and 
\begin{equation}
A_{1ij}^{T}=A_{2ij}=-\left[ \frac{\xi ^{\alpha }}{v_{k}}\frac{\partial v_{k}%
}{\partial \xi ^{\alpha }}\frac{\delta C_{ki}}{\delta \phi _{j}}+\frac{%
\delta C_{i}^{\alpha }(\phi )}{\delta \phi _{j}}\right] _{\phi
_{p}=v_{p},u_{i}=v_{i}} 
\end{equation}
in case of class II type of gauge fixing.
As a result, the Higgs pole masses of the multi-Higgs model are gauge independent in
both the class I and II type of the gauge fixing scheme.
\section{\bigskip DISCUSSIONS}
In this paper, we have investigated the gauge parameter dependence of the
Higgs pole masses in the multi-Higgs model in two different types of the gauge
fixing conditions. It turns out that the Higgs pole masses in case of the
multi-Higgs model when expressed in terms of the Lagrangian parameters,
are independent of the gauge parameter. Extension of this method in case of the
pole mass for the gauge bosons in the multi-Higgs model seems necessarily and 
this work is in progress.

\end{document}